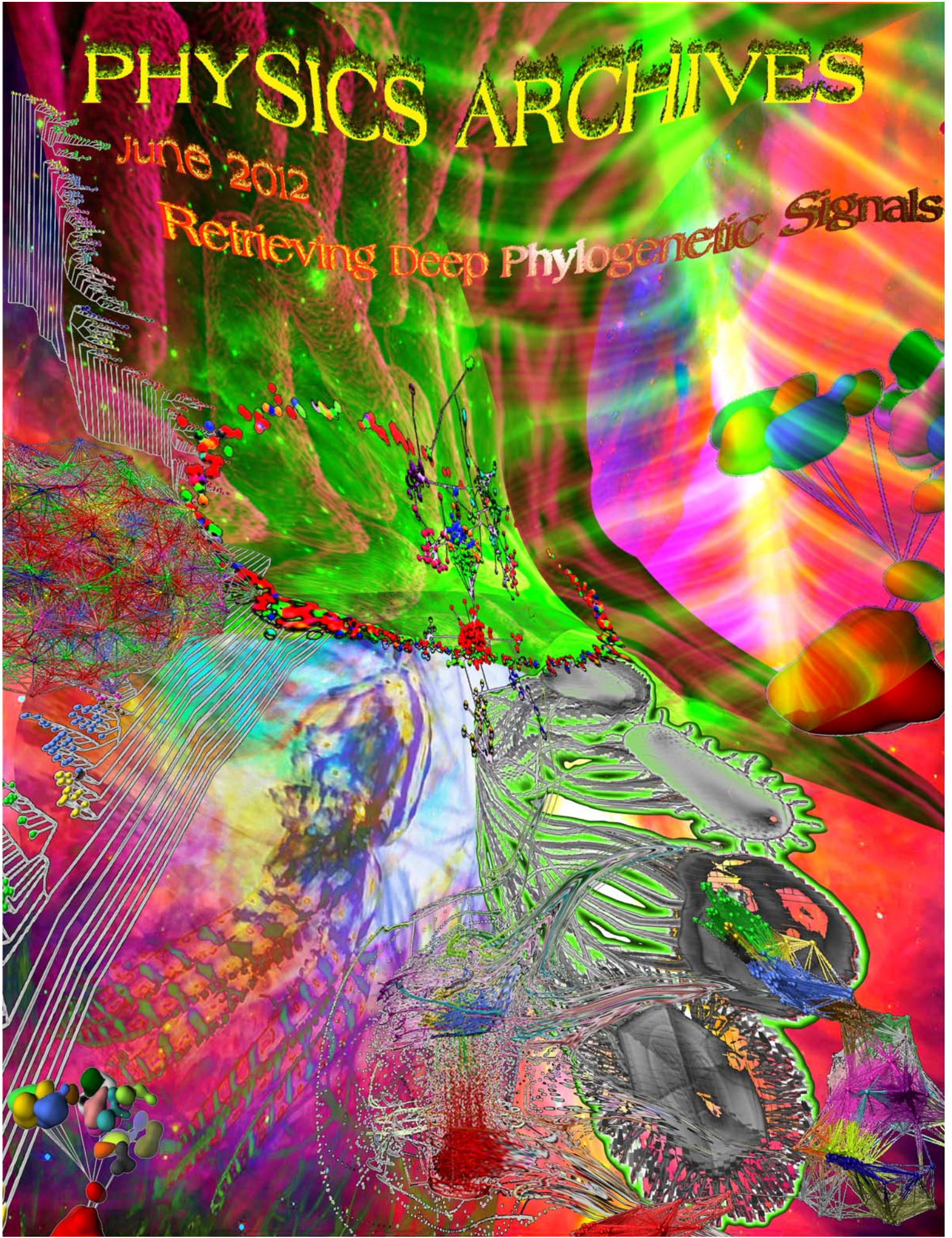

# Extraction of Deep Phylogenetic Signal and Improved Resolution of Evolutionary Events within the recA/RAD51 Phylogeny

Sree V. Chintapalli[1,2,3,*], Gaurav Bhardwaj[1,2,3,*], Jagadish Babu[4,5], Loukia Hadjiyianni[1], Yoojin Hong[4,6], Zhenhai Zhang[4], Xiaofan Zhou[5,7], Hong Ma[8], Andriy Anishkin[4], Damian B. van Rossum[4,5,#], Randen L. Patterson[1,2,3,#]

[1]Center for Translational Bioscience and Computing, University of California, Davis, USA

[2]Department of Biochemistry and Molecular Medicine, School of Medicine, University of California, Davis, USA

[3]Department of Physiology and Membrane Biology, School of Medicine, University of California, Davis, USA

[4]Center for Computational Proteomics, The Pennsylvania State University, USA

[5]Department of Biology, The Pennsylvania State University, USA

[6]Department of Computer Science and Engineering, The Pennsylvania State University, USA

[7]Intercollege Graduate Program in Cell and Developmental Biology, The Huck Institutes of the Life Sciences, The Pennsylvania State University, USA

[8]State Key Laboratory of Genetic Engineering, Institute of Plant Sciences, Center for Evolutionary Biology, School of Life Sciences, Fudan University, China

*These authors contributed equally to this work

#Address correspondence to:

**Randen L. Patterson**, 4220 Tupper Hall, One Shields Rd., Davis, CA 95616;

Tel: 001-530754-7660; Fax: 001-530-752-8520; E-mail: randen100@gmail.com

**Damian B. van Rossum**, 230-B Millennium Science Complex, University Park, PA 16802;

Tel: 001-814-863-0865; E-mail: dbv10@psu.edu

Running Title: Deep Origins of recA/RAD51 Recombinases

**Abstract**

**The recA/RAD51 gene family encodes a diverse set of recombinase proteins that effect homologous recombination, DNA-repair, and genome stability. The recA gene family is expressed in almost all species of Eubacteria, Archaea, and Eukaryotes, and even in some viruses. To date, efforts to resolve the deep evolutionary origins of this ancient protein family have been hindered, in part, by the high sequence divergence between families (i.e. ~30% identity between paralogous groups). Through (i) large taxon sampling, (ii) the use of a phylogenetic algorithm designed for measuring highly divergent paralogs, and (iii) novel Evolutionary Spatial Dynamics simulation and analytical tools, we obtained a robust, parsimonious and more refined phylogenetic history of the recA/RAD51 superfamily. Taken together, our model for the evolution of recA/RAD51 family provides a better understanding of ancient origin of recA proteins and multiple events leading to the diversification of recA homologs in eukaryotes, including the discovery of additional RAD51 sub-families.**

**Introduction**

recA recombinases are an ancient protein family that has evolved diverse roles in DNA management, including repair, recombination, and maintenance of genome stability [1-3]. There are three accepted subfamilies, namely: recA, RADα, and RADβ [4-8]. These families can be further subdivided into additional clades that have specific functions. For example, bacterial recA is a DNA-dependent ATPase that binds to single stranded DNA to promote homologous recombination; in eukaryotes, these functions are performed by RAD51 members [9-11]. Genetic knock-out of recA in bacteria leads to cell death due to the accumulation of deleterious mutations [12]. Similarly, RAD51 knock-out mice exhibit cell death and embryo inviability[13]. Further, DMC1, a eukaryote specific group, is required for meiotic recombination [14] with DMC1 knock-out mice manifesting truncated oogenesis. Thus, as a group recA/RAD51 proteins are of fundamental importance for cell-viability across all domains of life. Importantly, duplications of ancestral recA sequences and diversification of functions led to the increased complexity apparent in extant species [7,15].

Seminal phylogenetic studies on this superfamily by Lin *et al.* [16] proposed that: (i) bacteria contain only one recA gene, (ii) archaea contain two recA genes (RADA and RADB), (iii) yeast have four recA genes, and (iv) vertebrate animals and plants have at least seven recA genes [4,5,10,11]. This study obtained considerable support for orthologous groupings for recA, RADA, RADB, DMC1, RAD51, XRCC2, XRCC3, and RAD51B-D (see Figure 1A below for representation of their statistical inferences), and also postulated that eukaryotic recA genes evolved via two independent endosymbiotic transfer events. However, to obtain these groupings, many highly divergent sequences were omitted from the analysis because of their ambiguous branching pattern in the tree.

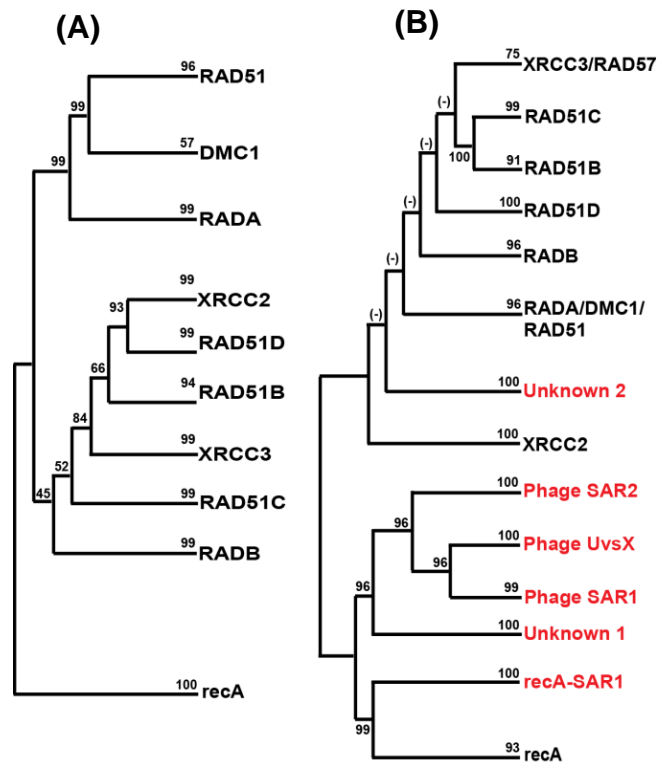

**Figure 1: Phylogenetic Inference of the recA/RAD51 Superfamily using MSA-based methods.** Representative phylogenetic trees of recA/RAD51 gene family as inferred in (A) Lin *et al.* (2006) and (B) Wu *et al.* (2011). Clades with metagenomic sequences that are unique to Wu *et al.* are demarcated in red. The notation (-) is indicative of no support for the given branching pattern.

More recently, Wu *et al.* [17] used a metagenomic survey approach and isolated a number of novel, and potentially ancient members of the recA family (i.e. recA-SAR1, Phage UvsX, Phage SAR1, Phage SAR2, Unknown 1, and Unknown 2). From this analysis, they concluded that: (i) these sequences are related to the recA/RAD51 protein family, (ii) several of these new groups are either viral lineages (e.g. bacteriophage) or archaea in origin, and (iii) one

| Groups | No. of Seq | Viruses | Meta-GOS | Bacteria | Archaea | Eukarya | Pairwise %Identity (ave in/btw group) |
|---|---|---|---|---|---|---|---|
| recA | 243 |  | ✓ | ✓ |  | Pr, Fu, Pl, | 61.5 \| 24.7 |
| RADA | 48 |  |  |  | ✓ |  | 56.8 \| 30.0 |
| RADB | 31 |  |  |  | ✓ |  | 44.0 \| 30.0 |
| RADAB | 5 |  |  |  | ✓ |  | 74.5 \| 30.0 |
| DMC1 | 55 |  |  |  |  | Pr, In, Nm, Fu, Pl, Ch | 59.2 \| 29.9 |
| RAD51 | 70 |  |  |  |  | Pr, In, Nm, Fu, Pl, Ch | 68.7 \| 29.6 |
| RAD51B | 15 |  |  |  |  | Pl, Ch (Pr) | 51.5 \| 30.0 |
| RAD51C | 24 |  |  |  |  | Pr, Pl, Ch | 51.4 \| 30.0 |
| RAD51D | 18 |  |  |  |  | Pl, Ch (Pr, Fu, In) | 48.7 \| 30.0 |
| XRCC2 | 15 |  |  |  |  | Pl, Ch (Pr, In) | 46.6 \| 30.0 |
| XRCC3 | 21 |  |  |  |  | Pl, Ch (Pr, In) | 48.9 \| 30.0 |
| recA-SAR1 | 10 |  | ✓ |  |  |  | 74.6 \| 30.0 |
| Phage SAR1 | 14 | ✓ | ✓ |  |  |  | 66.5 \| 30.0 |
| Phage SAR2 | 17 |  | ✓ |  |  |  | 73.3 \| 30.0 |
| Phage UvsX | 21 | ✓ | ✓ |  |  |  | 66.6 \| 30.0 |
| Unknown 1 | 6 |  | ✓ |  |  |  | 67.4 \| 30.0 |
| Unknown 2 | 20 |  | ✓ |  | ✓ |  | 57.1 \| 30.0 |

**Table 1: Qualitative and Quantitative Analysis of 17 Sub-Groups within the recA/RAD51 Superfamily.**
Abbreviations are as follows: Protists (Pr), Insects (In), Nematodes (Nm), Fungi (Fu), Plants (Pl), and Chordate (Ch). Parentheses in RAD51B, D and XRCC2, XRCC3 groups denote species which are putative members of the respective group but were not included in the phylogenetic inference because they disrupt the overall topology and can't be unambiguously assigned. These 14 sequences were subsequently added to the spatial dynamic clusters in Figure 9 as an indicator of the accuracy of their putative classification; they are as follows: {XRCC2_303290256_Micromonas_pusilla_Plants,XRCC2_332024988_Acromyrmex_echinatior_Insecta,XRCC2_255074101_Micromonas_Plants,XRCC2_66803939_Dictyostelium_discoideum_Protists,XRCC2_281210087_Polysphondylium_pallidum_Protists,RAD51D_170071670_Culex_quinquefasciatus_Insecta,RAD51D_321474080_Daphnia_pulex_Animal,RAD51D_111226459_Dictyostelium_discoideum_Protist,XRCC3_307191609_Harpegnathos_saltator_Insecta,XRCC3_281201100_Polysphondylium_pallidum_Protist,XRCC3_170044836_Culex_quinquefasciatus_Insecta,XRCC3_307171500_Camponotus_floridanus_Insecta,RAD51B_45685353_Chlamydomonas_reinhardtii_Protists,ID9_Unknown2_118195642_Cenarchaeum_symbiosum_Protists}

new group, designated Unknown 1, is very distant from the other groups and may be a fourth domain of life. Wu *et al.* [17] identified Unknown 1 as an environmental sequence with no useful information with respect to its sequence origin, which branches deeply (i.e. either between the three domains or as one of the deepest branches within a domain). Although these findings are potentially of great importance, the phylogenetic trees including these metagenomic sequences drastically differ from those of Lin *et al.* [16]. In particular, the branching pattern of archaea sequences, a key place in the history of recA recombinases, differs between these studies (compare Figure 1A and 1B above).

To discriminate between these two disparate phylogenetic results, we applied our recently developed Position Specific Scoring Matrix (PSSM)-driven algorithm, termed PHYlogenetic ReconstructioN (PHYRN), that is more accurate and robust for tree inference in highly divergent protein families[18]. PHYRN was benchmarked in simulated data sets with average pairwise identity <8.5% and was shown to be more accurate than Maximum Parsimony[19], Maximum Likelihood [20] and Bayesian [21] methods based on multiple sequence alignment. Importantly, this algorithm can handle large and diverse data sets, which may be required to discriminate between phylogenies proposed by Lin *et al.*[16] and Wu *et al.*[17]. We therefore applied PHYRN to study this important protein family.

The present study is subdivided into two major sections. The first describes PHYRN-based estimates of deep phylogenetic signal within the recA/RAD51 superfamily and compares the tree branching pattern, statistical support, and evolutionary inference by the distance-based Neighbor-Joining

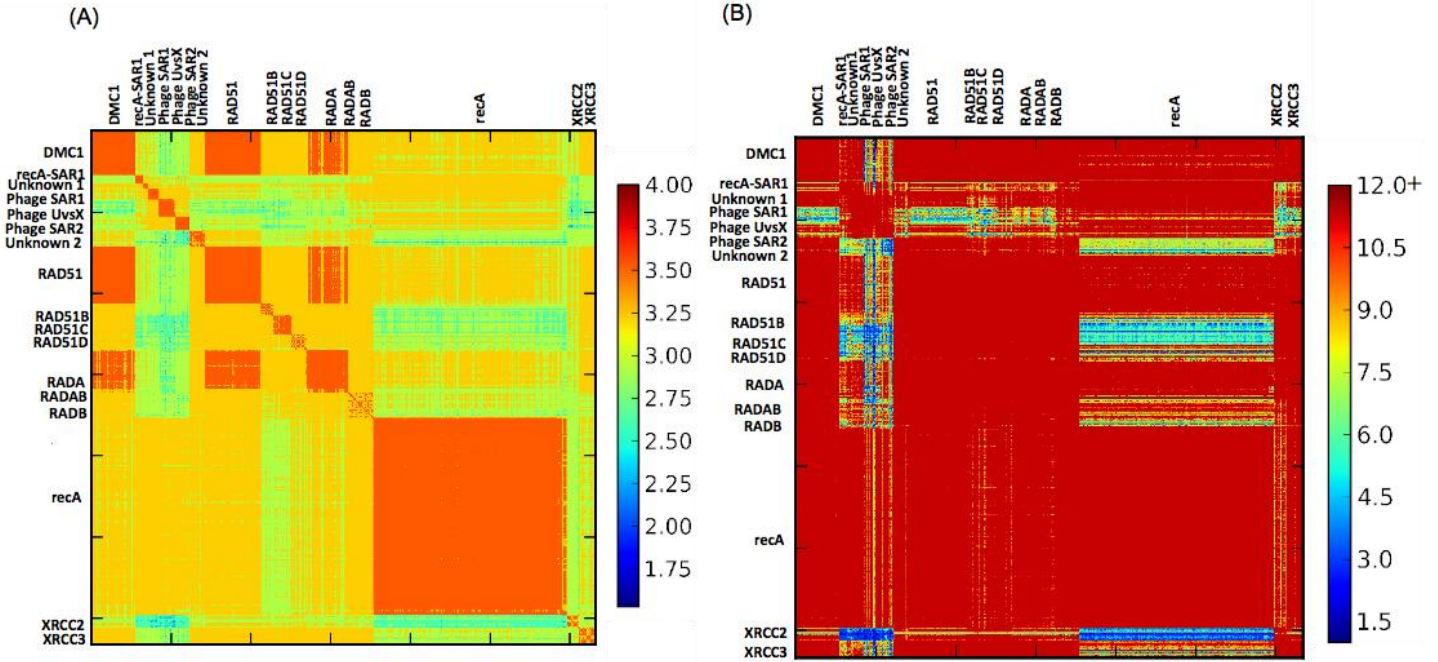

**Figure 2: Distribution and Characterization of PHYRN-Derived Phylogenetic Signal in recA/RAD51 Superfamily.** (A) Distribution of PHYRN Phylogenetic signal (% identity x % coverage) for recA/RAD51 superfamily. PHYRN score is calculated from alignments between full length query sequences and the respective recA/RAD51-specific PSSM library. PHYRN scores are represented as log-scaled values ranging from 0 (blue) to 4 (red). (B) Graphical representation of PHYRN phylogenetic signal of recA/RAD51 sequences (signal) as compared to their randomized versions (i.e. noise, 100 replicates). Comparative analysis is represented as log-scaled Difference Ratio (DR)

(NJ) algorithm for datasets representative of the Li et al.[16] and Wu et al. [17] studies. The second explores the application of Visual Molecular Dynamics software (VMD) for simulating, visualizing, and analyzing PHYRN-derived evolutionary distance estimates in a three-dimensional space representation, and dynamics of optimization in the spatial representation. We refer to this approach herein as "Evolutionary Spatial Dynamics" (ESD) simulations. While each is distinctive in its own right, these subsections are unified and integrated in their purpose: (i) to refine the phylogenetic history of recA/RAD51 recombinases, (ii) to showcase the utility of PHYRN for inference of deep evolutionary events in highly divergent datasets and (iii) to explore alternative scaling, scoring, and weighting procedures towards optimizing phylogenetic inference based on PHYRN measures. From the combined data, we propose a model of recA/RAD51 evolution that: (i) includes more diverse members of recA/RAD51 lineages and the new basal groups isolated by Wu et al. [17] from metagenomic sources, (ii) largely accords with the overall general pattern of Lin et al.[16], and (iii) lends support to the idea of the basal origin and diverse nature of metagenomic sequences as proposed by Wu et al.[17]. Taken together, our findings further resolve the deep origins of recA/RAD51 family and demonstrate the applicability/adaptability of PHYRN for phylogenetic inference of ancient protein families.

**Methods for PHYRN-based Measures, Tree Inference, and Statistical Analyses**

*Collection and Expansion of sequences-* 169 sequences used in Lin et al. [16] were collected and recA/RAD51 domain boundaries were defined using NCBI CDD at default settings[22]. Homologous regions thus defined were used as query set for expansion. PSI-BLAST[23] was used to collect homologous (recA/RAD51 domain containing) sequences from NCBI NR database with an e-value threshold of $1e^{-6}$ with 3 iterations. The top 10% scoring hits of expansion results from each sequence were retained. A data set comprised of 545 sequences was obtained using this approach. Further, we used PHYRN to align 195 metagenomic sequences from Wu et al. [17] against 545 recA-specific PSSM library. Based on the PHYRN composite score, these sequences were clustered using Pearson's Correlation and hierarchical clustering as available in Cluster 3.0[24]. Following,

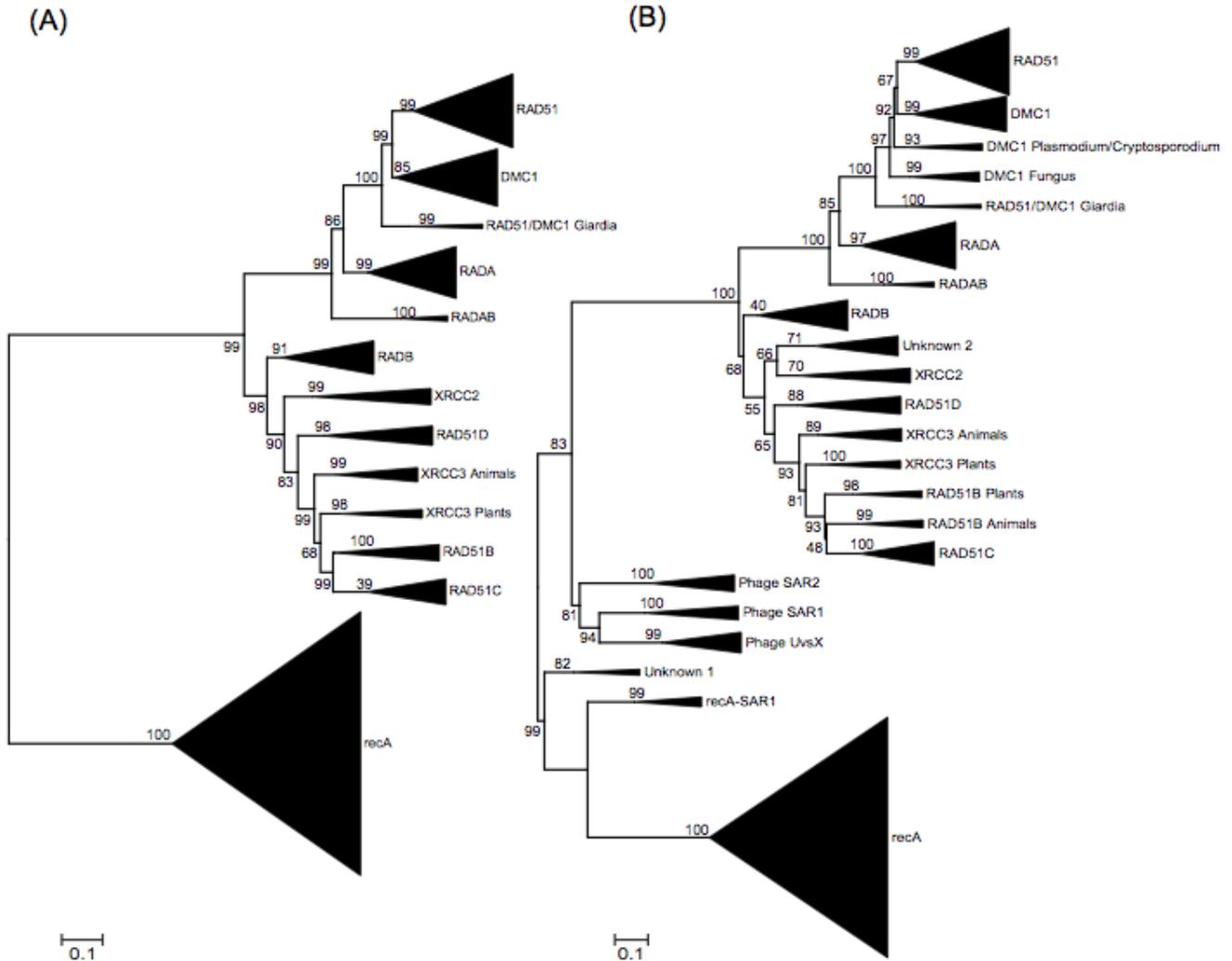

**Figure 3: Phylogenetic Inference of the recA/RAD51 Superfamily using PHYRN-NJ**
(A) Unrooted phylogenetic tree of Set-1 for 545 recA/RAD51 sequences. (B) Unrooted phylogenetic tree of Set-2 for 633 recA/RAD51 633 sequences (i.e. Set-2 comprises Set-1 +88 metagenomic sequences). Confidence values are calculated by jacknife resampling for 5000 replicates for both A-B. Scale bar is proportional to PHYRN-derived Euclidean distance scaled between 0-1.

88 sequences belonging to ID2 (PSAR1), ID5 (PSAR2), ID4 (PUvsX), ID15 (Unknown 1), ID 11 (RecA-SAR1) and ID9 (Unknown 2) clusters were added into the previously described 545-sequence data set. For clarity and transparency, the sequence distribution of Set-1 and Set-2 reported above as well as orthologous and paralogous pairwise comparisons reported in Table 1 are in the absence of 14 sequences which were removed during dataset curation due to their disruption of cladistic separation in subsampled trees and their unambiguous classification by phylogenetic analyses. These sequences are reported in Table 1 Legend and utilized in classification vs. randomization tests described in Subsection II.

***PHYlogenetic ReconstructioN (PHYRN)***- The pipeline for the PHYRN algorithm is described in detail elsewhere [18]. Briefly, the input is a set N of amino acid sequences and set M of their associated position specific scoring matrices (PSSM). The output is a tree *T*, leaf-labeled by the set N. PHYRN uses a custom code for rps-BLAST for recording positive alignments between simulated sequences and their respective PSSM library. For a given profile M, the matrix is populated 0 for no alignment or as a positive product score for the alignment with best PHYRN score (%identity X %coverage) retrieved with an e-value threshold of $10^{10}$.

Equation 1-2, Percentage identity (%i) and percentage coverage (%c) is defined as follows:

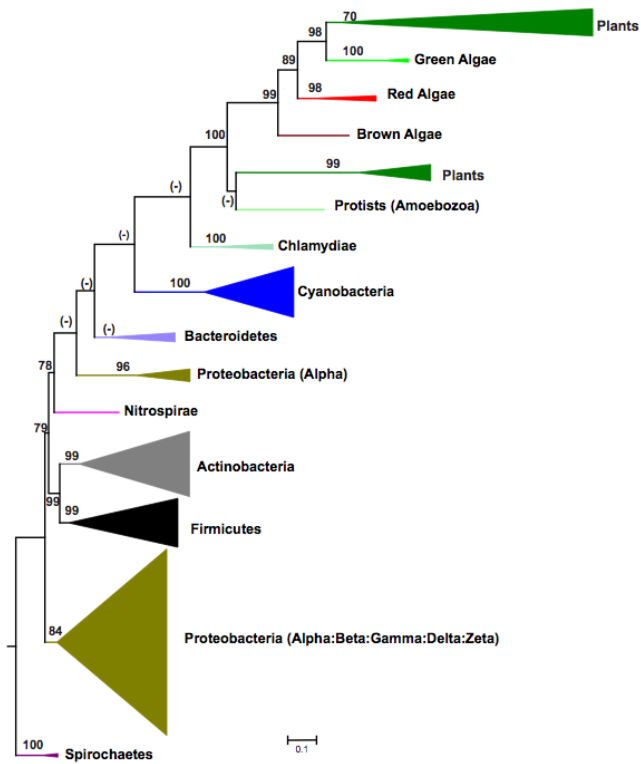

**Figure 4: Evolution of recA sequences**
A phylogenetic tree of 242 recA sequences inferred using PHYRN-NJ and rooted with Spirochaetes. Branch statistics are derived from jacknife resampling tests for 5000 replicates. The notation (-) is indicative of no support for the given branching pattern. Scale bar is proportional to PHYRN-derived Euclidean distance scaled between 0-1.

Eq. 1: %i = [(Number of Identical residues in alignment) / (Alignment length including gaps)]

Eq. 2: %c = [(Alignment length in query excluding gaps) / (Sequence length of PSSM)]

Mathematical derivations show that this PHYRN product score is equivalent to [(1-(Alignment restricted p-distance))*(1-PHYRN gap-weight)]. Alignment Restricted p-distance ($p_{ARP}$) is defined as proportion of amino acid sites different in alignment defined as a function of PSSM length. It is calculated by dividing the number of non-identical amino acid sites by total length of the PSSM. PHYRN Gap Weight ($\omega_g$) is defined as proportion of gaps defined as a function of alignment length. It is calculated by dividing total number of gaps in alignment by length of alignment. From the NXM matrix, PHYRN calculates the Euclidian distance between each query, which can then be depicted as a phylogenetic tree using a variety of distance-based tree-building algorithms (e.g. Neighbor-Joining [25], Minimum Evolution[26], FASTME[27] etc.).

***Implementation of PHYRN for recA/RAD51 sequences-*** The recA/RAD51 domain boundaries were defined in the full-length sequences using NCBI CDD with default settings [22]. These homologous regions were extracted using a custom python script and used to generate recA-specific PSSM library- using codes provided in PHYRN v1.6 package (http://code.google.com/p/phyrn/). To increase the specificity of the PSSM library, we first collected all putative recA/RAD51 containing proteins, and subsequently used these sequences as a target database for pssmgen script in the PHYRNv1.6 package. Previous results with PHYRN have shown that e-value of $1e^{-6}$ provides the best results with the non-redundant (NR) NCBI database [18]. Since our target recA/RAD51 database is significantly smaller in size, and the e-value threshold scales proportional to the size of target database, we used an e-value of $7e^{-13}$ for PSSM generation. In the next step, full-length sequences were aligned with this PSSM library, and alignments were encoded in a composite score matrix. While running rpsBLAST, we used a "–b" value setting that shows alignments for only the top scoring 75% of total PSSMs. In experiments with ROSE-derived synthetic protein families we validated that "–b" equal to 75% of total PSSMs provides the most accurate results (data not shown). This composite score matrix was further used to calculate a Euclidean distance matrix. Neighbor-Joining (NJ) algorithm as implemented in MEGA v5.03 [28] was used to calculate phylogenetic trees from the Euclidean distance matrix.

***Implementation of MSA/Protdist/ML-*** Optimal multiple sequence alignment (MSA) was calculated using MUSCLE v3.8 [29] with default settings. Protdist from PHYLIP package v3.69[30,31] was used to calculate evolutionary distances. We used MEGA v5.03 to calculate the best protein substitution model for distance calculation. Based on these calculations, we used protdist with JTT (Jones, Taylor and Thornton)[32] as a substitution matrix of choice, and gamma correction value of 0.8. For maximum likelihood (ML) trees, we used RAxML v7.2.8 [20] with MUSCLE alignment as input. RAxML was used with JTT as substitution matrix of choice. Empirical frequencies were estimated from the data in hand (+E setting), and a gamma correction value 0.8 was used. All other settings were used as defaults.

***Statistical Resampling-*** Statistical support for PHYRN was calculated using jacknife resampling, while for protdist and ML trees we used Bootstrap resampling. For jacknife resampling of PHYRN data, 80% of data points were randomly subsampled without replacement from the PHYRN NXM matrix. 5000 random replicates were generated in this manner and Neighbor program from PHYLIP package [30,31] was then used to calculate Neighbor-Joining trees. The Consense program from PHYLIP package[30,31] was used with majority rule consensus method to calculate a consensus tree of 5000 replicates; these isometric consensus trees are shown in collapsed version (Figure 3, 4 & 5) and full extended trees are provided upon request. Confidence values thus obtained were compared for three-points of reference in the PHYRN trees and where appropriate were appended to branch labels in our PHYRN trees (Figures 3-4). The symbol (-) denotes an unsupported branch in the tree. For protdist and ML method, bootstrap resampling was conducted using their default settings with 1000 and 100 replicates respectively (Figure 5).

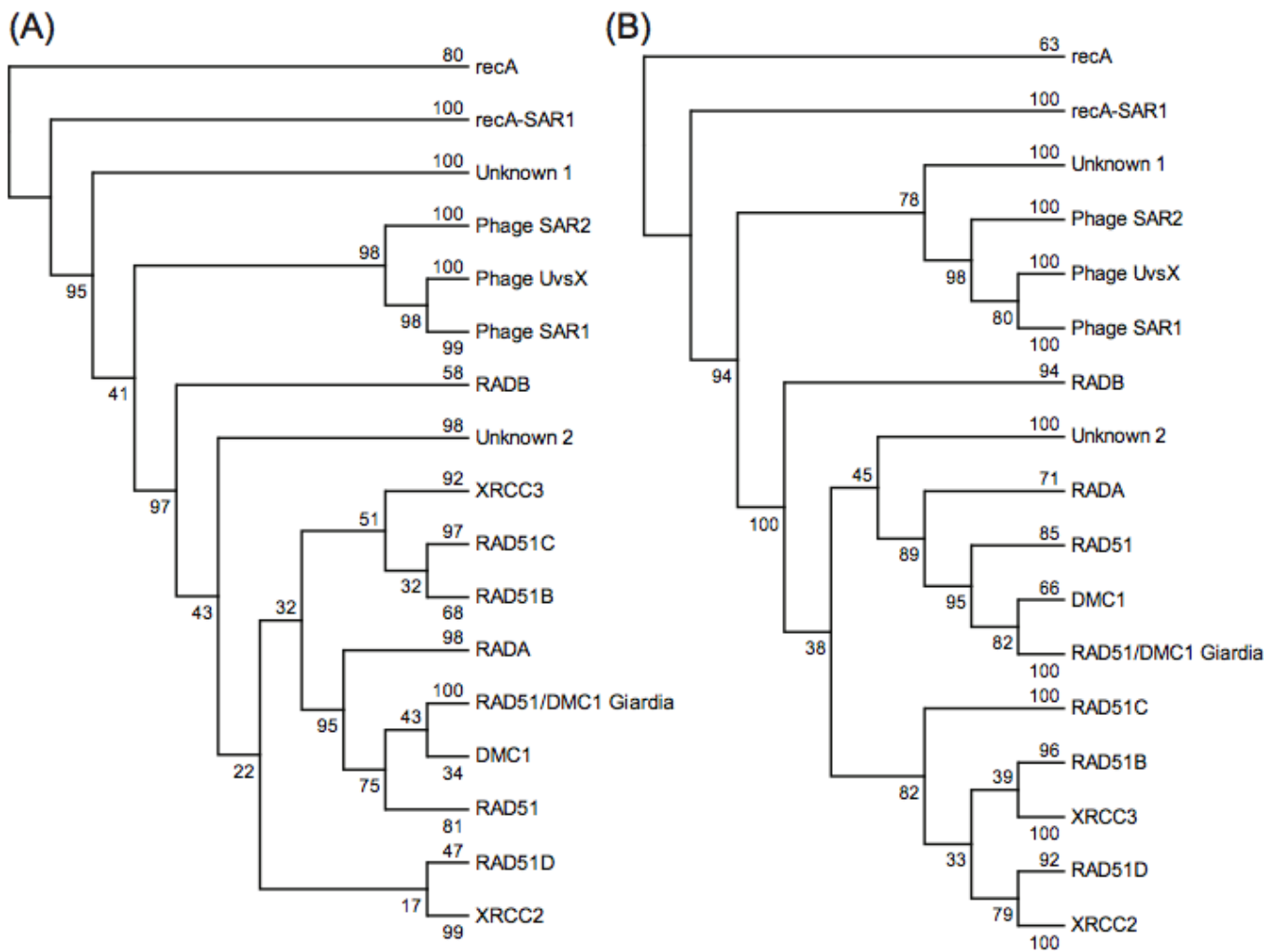

**Figure 5: Phylogenetic Inference of recA/RAD51 protein family inferred using MSA-based approaches.** Consensus phylogenetic trees of 633 recA/RAD51 sequences inferred from an MSA obtained using MUSCLE. (A) For NJ tree, Protdist from PHYLIP v 3.9 was used to calculate distance matrix with JTT as substitution matrix of choice, and gamma value of 0.8. (B) For RAxML, trees were calculated with JTT as substitution matrix of choice, and empirical frequencies and gamma value were calculated from the data set at hand. In A-B, confidence values were calculated using bootstrap resampling method with 1000 or 100 replicates, respectively.

*Randomization Test for PHYRN-Derived Phylogenetic Signal-* We conducted a randomization test to quantify a signal-to-noise ratio in our measurements of phylogenetic signal. In this test, each full-length query sequence was randomized in its linear order of amino acids without replacement. Randomized sequences were then aligned with our recA-specific PSSM library and alignment scores were encoded in a new NXM-random data matrix. This randomization step was repeated for 100 different random replicates and an average and standard deviation for each coordinate was recorded. A Difference Ratio (DR) was calculated for each coordinate using the following equation and represented as log-scaled values:

Equation 3: Difference Ratio = (composite score$_{wt}$-average composite score$_{random}$)/SD$_{random}$.

## Results

### Section I
### Construction of recA/RAD51 Data sets

Our initial data set comprised 169 sequences that were obtained from Lin et al.[16]; this data set was expanded in number and diversity using PSI-BLAST[23] against the non-redundant NR NCBI database (see Methods). After this expansion, we obtained 545 sequences, denoted as Set-1. To obtain direct comparisons with the Wu et al. [17] study, we included 88 metagenomic sequences isolated from the Sorcerer II Global Ocean Sampling Expedition (GOS), termed here Set-2. In Table 1, we present qualitative and quantitative statistics for both data sets, including the number and distribution of sequences in each sub-group of the recA/RAD51 family. For groups with sequences representative of eukaryotic lineages we have further annotated the sequence diversity to demarcate the presence of protist, insect, nematode, fungi, plant, and/or chordate species. Phage SAR1, Phage SAR2 and Phage UvsX are enterobacteriophage sequences. We identified archaea specific group, RADAB, which shows a split recombinase recA domain with the presence of a large insertion. With respect to sequence similarity, Set-1 and Set-2 are conserved within orthologous groups, but divergent between paralogous groups (~30% average pairwise identity between groups as measured by MUSCLE [29], see Table 1). All sequences utilized in this study as well as the chopped boundaries utilized for PSSM generation are available upon request.

### Quantification of Phylogenetic Signal within the recA/RAD51 Superfamily

Since all sequences in Set-1 and Set-2 share a common recA domain, these homologous domains were used to construct a recA/RAD51 specific PSSM library (see [18] and Methods for complete description of PHYRN implementation). Subsequently, full-length sequences from each dataset were aligned with their respective recA/RAD51 PSSM library. The results from these alignments were collected and the alignment statistics (i.e. composite score=percentage identity X percentage coverage) were encoded as a N-query by M-PSSM (NXM) similarity matrix. The heat map in Figure 2A graphically represents the phylogenetic signal of the NXM matrix for Set-2 represented on a log scale (red = maximal possible score, 10,000; dark blue = lowest possible score, 0). These data suggest that all sub-families have excellent signal within group, and a varying amount of signal across paralogous sub-families.

To further quantify the signal-to-noise ratio we conducted a randomization test, in which each full length query sequence was randomized in its linear order of amino acids, without replacement, so that it retained the same length and amino acid composition. Randomized sequences were then aligned with the respective wild-type recA-specific PSSM library and alignment scores were encoded in a new NXM-random data matrix. This process was repeated for 100 different random replicates and an average and standard deviation for each coordinate was recorded. A Difference Ratio (DR) was calculated for each coordinate using Equation 3 (see Methods). Hence the DR is a reflection of the amount of signal above background inherent to each comparison. The DR is plotted as a heat map in Figure 2B represented on a log scale (blue=lowest SD above random, red=largest SD above random). Overall there is a strong signal-to-noise ratio across all the groups. Notably, metagenomic sequences also show strong phylogenetic signal against other groups, thereby justifying their inclusion in this phylogenetic study.

### Phylogenetic Inference of the recA/RAD51 Family

Unrooted phylogenetic trees for both Sets (Figure 3A and 3B, respectively) were constructed from a Euclidian Distance of the NXM similarity matrix to produce a NXN distance matrix. Subsequently, a phylogenetic tree was inferred by distanced-based NJ algorithm as described previously [25]. In the tree of Set-1, we observe three major clades, namely: (i) recA (ii) RADα and (iii) RADβ (see Figure 3A). Upon close inspection, the branching pattern is largely in accordance with Lin et al.[16]; however, there are some notable differences. Specifically: (i) we identified a new achaeal group, RADAB, which is in an intermediate position between RADA and RADB archaea groups, (ii) we were able to include more representatives from protist, insect, nematode, archaea and bacterial sources across different clades, and (iii) our tree displays more robust statistical support across deep branches.

Between both Sets, we also observe distinctive branching points at several positions. In the PHYRN-NJ tree of Set-1, ancestral RAD51/DMC1 Giardia sequences are outgroups to both DMC1 and RAD51 (DMC1 and RAD51 were monophyletic in Lin et al.). The presence of both DMC1 and RAD51 members in Plasmodium (chromalveolate) suggests that duplication events leading to the origins of DMC1 from a common ancestor of DMC1 and RAD51 likely happened after the evolution of alveolates (i.e. "with cavities", a major line of protists). In the PHYRN-NJ tree of Set-2, fungal sequences seem to be misplaced, as there are ascomycetes (i.e. commonly called "sac fungi" or "cup fungi" for the cup-shaped fruiting bodies) both before and after the alveolates. Conversely, the PHYRN-NJ tree from Set-1 shows a clear demarcation of DMC1-fungal and RAD51-fungal sequences. It is possible that the addition of metagenomic sequences may have led to a decreased resolution of these specific groups. Another difference between PHYRN-based inferences of Set-2 is that XRCC2 occupies a phylogenetic position closer to the archaeal ancestors with high statistical support. Finally, XRCC3 forms a paraphyletic group (i.e. metazoans [animals] outgroup viridaeplantae [green plants] members). This could be either due to a PHYRN-NJ branching error or a differential evolutionary rate between plants and animals.

Wu *et al.*[17] identified several new putative members of recA/RAD51 sequences from metagenomic sources. It is possible that the inclusion of these sequences would further refine our understanding of the deep origin of recA/RAD51 family. Indeed, inclusion of the metagenomic sequences (Figure 3B) leads to topological and statistical changes compared to the tree inferred for Set-1 (compare Figure 3A vs. Figure 3B). Interestingly, the metagenomic groups occupy divergent positions in the tree; in fact, Unknown 1 attains the most basal position in our PHYRN-NJ tree. Our present study supports a scenario in which RAD$\alpha$ and RAD$\beta$ share a common ancestor. This inference: (i) is a more parsimonious scenario assuming a recA/Unknown 1 root, (ii) accords with Lin *et al.*[16], and (iii) is in contrast to the tree in Wu *et al.*[17].

We also observe that endosymbiotic transfer events from bacterial recAs also contributed to the evolution of eukaryotic recA proteins (Figure 4). Specifically, multiple gene transfer events from cyanobacteria and chlamydiae (i.e. obligate intracellular pathogens AKA 'energy parasites' of bacterial phylum and class) led to evolution of chloroplast recAs in accordance with the literature on the origins of chloroplast[33]. We also observe another clade of viridaeplantae members that shows closer relationships with protist members. These recA sequences are nuclear in location, and may represent nuclear localized copies of endosymbiotic DNA, or products of secondary or tertiary endosymbiosis involving protist members. Moreover, our study infers that Gram positive bacteria (Actinobacteria and Firmicutes) form sister taxa in rooted trees.

Finally, we compared the PHYRN-NJ tree shown in Figure 3B to phylogenies inferred using multiple sequence alignment-based methods (Figure 5). Notably, both Muscle-NJ and Muscle-RaxML trees show similar positioning of metagenomic groups as compared to PHYRN-NJ; however, the Muscle-NJ tree shows a less robust statistical support compared to Muscle-RaxML and PHYRN-NJ trees. Importantly, the Muscle-RaxML tree predicts a non-parsimonious branching pattern for RAD$\alpha$ and RAD$\beta$. Moreover, in this tree RAD51 Giardia sequences would have evolved after the emergence of more complex mammalian DMC1 & RAD51 members.

Taken together, PHYRN-NJ provides a more refined, statistically robust, and logical phylogenetic inference for these data. However, even the PHYRN-NJ tree lacks resolution at some nodes, specifically for the events after the emergence of Unknown 2 (archaea) and before the diversification of RAD51 groups (XRCC2, XRCC3, RAD51B-D). Hence, the inclusion of metagenomic sequences leads to a loss of resolution and robustness with respect to the DMC1 and RAD51B lineages. Also, in the PHYRN-NJ tree, there are some likely topological errors, such as the position of fungal DMC1 sequences, even though it receives strong statistical support in the resampling analysis. These types of errors might be a function of: (i) missing sequences in the metagenomic groups, (ii) missing protists, nematodes, fungi, and insect sequences in higher-order groups that we could not find or could not include in the tree (see Table 1), (iii) possible sequencing errors for some representatives, (iv) branching errors by NJ, and/or (v) inaccurate distance estimates by PHYRN for some sequences.

It is important to reiterate that PHYRN is PSSM-based but MSA-independent. In other words, PHYRN utilizes a MSA in the construction of individual PSSM libraries but utilizes sequence-profile pairwise comparisons for alignment statistics. Tree inference is made from these statistics in matrix format not from analysis of a MSA of the various sequence alignments. Traditional phylogenetic methods rely on an MSA for refinements to tree inference (e.g. gamma parameters, differential models of substitution patterns, pairwise/complete deletion of gaps). Therefore, we sought a complementary algorithmic approach that: (i) explores different scoring, scaling, and weighting functions that could be applicable to PHRYN-based measures and (ii) offers added value to the visualization/analytics of tree inference of large datasets based on these measures.

**Section II**

**Rationale for Application of Evolutionary Spatial Dynamics to Sequence Similarity**

As foreshadowed above, a key and general challenge for evolutionary inference is the ability to analyze and visualize the relationships between large numbers (hundreds-many thousands) of sequences. This is especially relevant in datasets of high sequence divergence and/or ancient origins- large phylogenies can refine/resolve evolutionary inference that are not adequately handled by smaller phylogenies [34-39]. Moreover and as noted, a general problem for PHYRN-based measures is the absence of an MSA from which to explore refinement strategies outside of PSSM generation.

However, there is already an established field of study with sophisticated and high-performance parallel algorithms/visual applications that is able to handle millions of linked data points- software for visualization of molecular dynamics simulations. We have used Visual Molecular Dynamics (VMD) software[40] herein for dynamic spatial representation and analysis of evolutionary similarity data in 3D space. We theorize that this approach can be adapted to study protein sequences in new and powerful ways. Indeed, there are numerous reports on the concept of 2D/3D visualization of sequence similarity networks and applications to adjacent techniques (e.g. 3D visualization of gene expression data) [41-43]. The recent study by Babbitt and colleagues [43] is particularly notable and encouraging. Their multiple networks, each derived from evolutionarily conserved proteins families, were interactive, intuitive and aesthetically pleasing; offering the benefit of overlaying/visualizing orthogonal data inherent to the functional protein characteristics. In addition, these authors presented a compelling quantitative argument that sequence similarity networks can competently depict sequence similarity relationships (i.e. the mathematics underlying the conversion of one-dimensional evolutionary distance to higher dimensional (2D and 3D) spaces is a satisfactory depiction). Their claim is that, "*The most valuable feature of sequence similarity networks is not the optimal or most accurate display of sequence similarity, but rather the flexible visualization of many alternate protein attributes for all or nearly all sequences in a superfamily*"[43].

Key points of consideration for the present study is that when these authors decreased the stringency of their statistics, allowing an increasing number of low-identity alignments into their networks, these networks lost resolution and largely collapsed. These authors also stressed that, "*While there are many similarities between the interpretations that can* be made from the information provided in a network and a tree, phylogenetic trees are based on an explicit evolutionary model that is missing from sequence similarity networks. Thus, networks are not an adequate alternative to a tree, as the interrelationships they depict cannot be used as a basis for inferring evolutionary history. Indeed, there is a fundamental difference between the network composed of nodes representing contemporary protein sequences that may be connected with cycles, and the acyclic Steiner tree with introduced ancestral nodes that can be used to describe a phylogenetic tree". We agree that networks are not a standalone alternative to phylogenetic trees; however, it is reasonable to consider that they could be utilized synergistically in complementary ways.

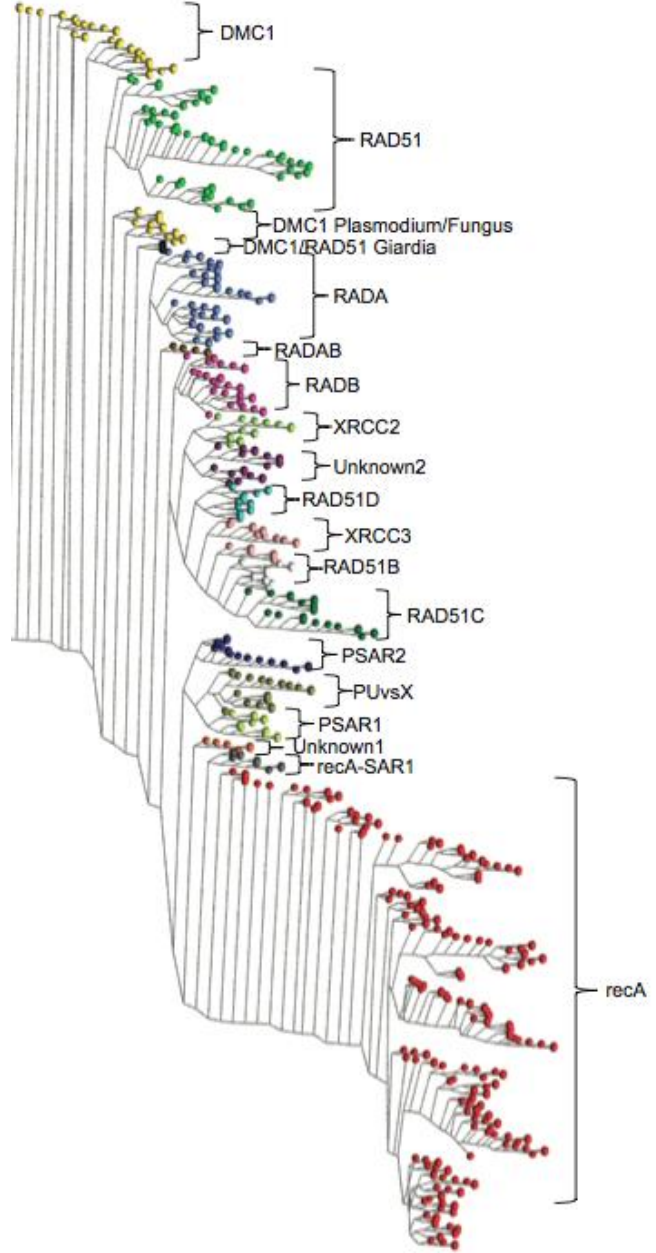

**Figure 6: Conventional 2D representation of the PHYRN tree using VMD graphics.**
Conventional graphical representation of the PHYRN-NJ derived tree for 633 recA/RAD51 sequences represented in 2D space and colored by subtype.

Overall, these types of network approaches and perspectives are quite promising and offer broad utility; however, these approaches have not yet become widely used or accepted. This may in part be due to a lack of convenient/quick and universal implementation of 3D visualization (i.e. it always involved considerable effort), and/or most approaches end at the visualization stage and do not bring any critical new information unattainable by other means. We hypothesize: (i) that Evolutionary Spatial Dynamics (ESD)-based approaches may bring added value to evolutionary studies in ways not reflected previously, and (ii) that spatial dynamic clustering, in the context of PHYRN-derived branching patterns, can be used as complementary basis for inferring evolutionary history. Indeed, a specific benefit of molecular visualization software is the convenient display of time-varying positions of a dynamic network of nodes. This in turn might allow the user to monitor and infer the finer details of inter group connectivity, and/or phylogenetic signal. As a step towards a complementary way of simulating, visualizing, and analyzing PHYRN-derived phylogenetic signal, we explored the utility of VMD for representing spatial dynamics for all- against-all sequence similarity.

VMD software (v1.9.1), developed by the Theoretical and Computational Biophysics Group in the Beckman Institute for Advanced Science and Technology at the University of Illinois at Urbana-Champaign[40], provides a rich set of graphical representations with Tcl scripting capabilities; this allows for a flexible interface for custom-written routines for visualization, transformation and analysis of structures. We developed an in-house VMD-supported algorithm to simulate and visualize the spatial dynamics of protein sequences based on PHYRN-derived estimates of sequence similarity and evolutionary distances. In theory, this same approach to modeling evolutionary distances could be applied to other types of matrix-based data (e.g. population, sociological, epidemiological, gene expression data) and would offer several distinct advantages that may not be available with other software.

Added value to this VMD-supported approach may include: (i) 2D/3D trees may give much more freedom for positioning branches than traditional methods allow and lead to increases in visualization and/or pattern recognition especially for large datasets; (ii) There are numerous ways to visualize/render different tree and/or protein characteristics such as the size and color of the nodes and connections, density maps in space visualizing clusters of nodes, and/or protein structural/functional data (iii) It is all scriptable - commands for positioning nodes, drawing/removing labels and shapes in space, aligning two trees, comparing branches using RMSD are easily implemented; (iv) Convenient batch-selection commands- for a mega-tree, clicking a few thousand times is unthinkable; however, VMD allows selection by numerous criteria, both in commands and menu, such as all the relatives in certain vicinity. In addition, one can use additional PDB fields (radius, charge, segment name, residue name, beta, occupancy and user fields) for storing/managing additional information; (v) The clustering procedure can be animated to show the "growth of the evolutionary tree"; moreover, this animation (with exploration of the tree, showing part by part or rotation) can be exported into AVI movie for dissemination or exploration of the tree showing rotation or different time steps; (vi) The tree can be exported into VRML format for viewing in any 3D viewer (including online plugins on the internet for exploration); (vii) VMD can provide stereo views in numerous formats which may be useful for 3D trees; and (viii) VMD can provide output for a 3D printer, so the tree can be actually molded in colored plastic.

**Implementation of Evolutionary Spatial Dynamics to Evolutionary Measures**

For adequate application of molecular visualization tools for simulation/visualization of evolutionary data, it is necessary to have a consistent way to position the nodes in space (each node reflects a full length protein sequence). Our approach is based on pairwise proximity in 3D space reflecting the pairwise sequence similarity or pairwise "evolutionary distance" estimated by PHYRN. The input files for the spatial dynamic algorithm are as follows: (i) the PHYRN-derived NXM data matrix and (ii) an XML file of the PHYRN-NJ derived tree. Importantly, the branching order inferred by PHYRN-NJ does not inform the spatial dynamic algorithm (i.e. spatial positioning is based only on the pairwise distances); however, this XML file is read for the sequences list and naming convention of the tree leaves, and can be overlaid onto the spatial distribution, and rendered as lines for clarity and cross correlation of nearest neighbor estimates during the annealing process and inspection of the results. From a visual perspective, the more untangled these lines are, the more consistent the PHYRN-NJ branches compare with the respective spatially minimized tree structure.

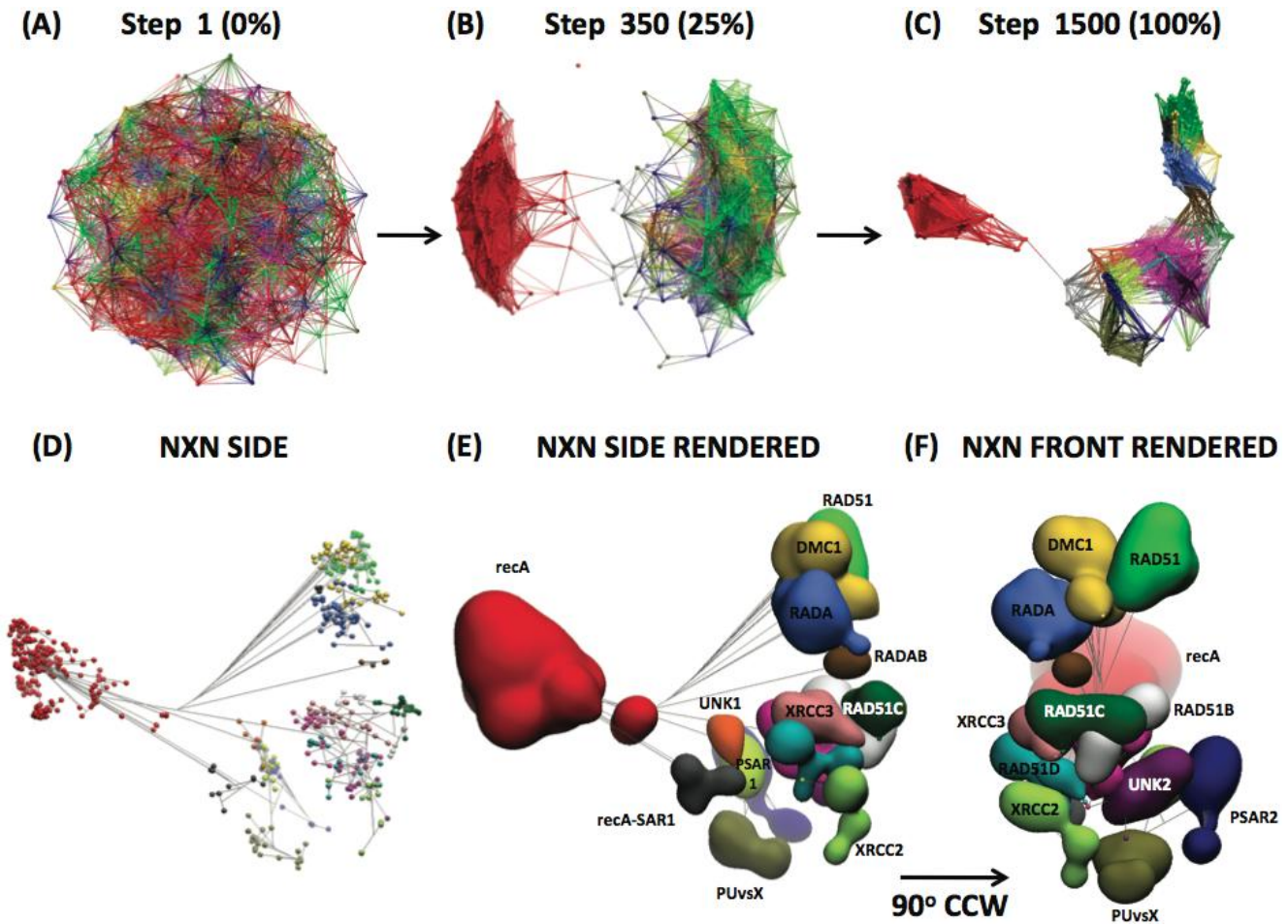

**Figure 7: Progression of a Typical NXN Evolutionary Spatial Dynamic Simulation.**
In this simulation of 633 recA/RAD51 sequences each sequence is represented by a node in space, and colored by their respective group. Each node is connected by harmonic bonds to all other nodes. For clarity, snapshots in A-C are rendered to show connectivity to close neighbors only and individual recA/RAD51 subtypes are color coded. (A) The simulation begins with all points randomly distributed in the simulation cell; a point is chosen at random to begin the distance optimization process. (B-C) For each cycle, each point is repositioned in space to minimize the deviation for the PHYRN-derived target pairwise distances to all other points (i.e. relocation towards the position of minimal force from all bonds leading to the current node). This process iterates by a random-optimization-order, wherein every cycle we randomize the order in which we take nodes for individual optimization, and the algorithm proceeds until the tree structure converges to equilibrium (panel C is a snapshot after 1500 iterations). (D) Final spatial positioning of all the nodes after the NXN spatial simulation reaches stationarity. (E-F) Side and Front cloud rendering views of the final node positioning in the NXN spatial simulation, respectively. Nodes belonging to the same clade are rendered in their respective clouds.

Alternatively, a two-dimensional VMD-supported version of the PHYRN-NJ tree can be output, rendered in various ways, rotated in 360°, and interrogated in an interactive manner. Figure 6 depicts this tree output; 633 full length recA/RAD51 sequences are represented as nodes in space and colored their respective group. Notably, there is a critical difference between ESD-based trees (see Figures 7-10) and the conventional tree shown here. Specifically, in ESD-based trees the direct distance between any two sequences is, on average, proportional to their sequence similarity by the PHYRN-derived metric, while in the conventional tree representation the evolutionary distance between any two sequences is defined by a "return-trip" from one sequence towards the nearest split point and back to another sequence. Along this route, only distances in one dimension (i.e. horizontal x axis) are summed up for the total distance/similarity. Therefore, sequences positioned far away from each other in the conventional tree (i.e. along the orthogonal y axis) might still be closely related and could be visually misleading. In contrast, within ESD-based trees the spatial proximity between nodes always reflects evolutionary relationships. Another distinction is that in the conventional representation, any branch can be "flipped" around the split point in the y dimension without any change in the presented information (i.e. any combination of flips would be equivalent). However, in ESD-based tree the 3D spatial position of nodes is based on optimized all-to-all relationships, therefore their position in space is meaningful and non-random, and cannot be changed arbitrary.

For conversion from NXM PSSM matrix data ($M_{ij}$, ranging from 0 (no similarity) to 10000 (identical)) to spatial pairwise "target'" distances $T_{ij}$ (i.e. the set of pairwise distances that we are attempting to position in 3D space) we use the following formulas:

Equation 1: Conversion from NXM matrix data ($M_{ij}$) to spatial pairwise "target'" distances $T_{ij}$:
Eq. 1a: Linear transformation

$$T_{ij} = 10000 - \min(9999, M_{ij})$$

Eq. 1b: Logarithmic transformation

$$T_{ij} = \ln\left(\frac{\min(9999, \max(1, M_{ij}))}{10000}\right)$$

Equation 1a preserves the linear correspondence between the distances measured in 3D. However, for sequences with low similarity a significant change in similarity level will translate only into a small difference in spatial positioning. For a pair of sequences that have four-fold difference in similarity to a third sequence (e.g. 10 and 40, 400% increase), their spatial distance to that third sequence will differ only marginally (i.e. 9990 and 9960 respectively, 0.3% decrease). We reasoned that better relative positioning of the nodes with low similarity may be achieved with the use of the log transformation (i.e. Equation 1b). In this case, sequence similarity $M_{ij}$ decreases exponentially with 3D spatial distance, $T_{ij}$. In our preliminary simulations, both linear and logarithmic transformation lead to a similar minimized spatial representation for the NXM matrix (data not shown), while the latter transformation leads to more cladistic separation of leaves (i.e. less mixing).

If one considers similarity to a specific PSSM library $M_{ij}$ as a coordinate of the node $i$ along the axis $j$ in multidimensional PSSM space, a Euclidian distance and linear conversion to spatial scale between two points can be defined by Equation 2 and 3, respectively:

Eq. 2: $N_{ij} = \sum_{k=1}^{m} \frac{(\vec{R}_j - \vec{R}_i)(|\vec{R}_j - \vec{R}_i| - T_{ij})}{|\vec{R}_j - \vec{R}_i|}$

Eq. 3: Conversion from NXN matrix data to spatial pairwise distances:

$$T_{ij} = \frac{\max(10, N_{ij})}{1000}$$ Linear transformation

In this paradigm, each spatial node is "bonded" to all other nodes. The ideal target length of a pairwise bond is the PHYRN-derived distance between the specific nodes, which is then transformed to a spatial scale. The algorithm begins with all nodes randomly distributed in the simulation cell and iterates by a random-optimization-order, wherein every cycle we randomize the order in which nodes are taken for individual optimization. The algorithm proceeds until the tree structure converges to some equilibrium shape (i.e. stops moving-nodes find a minimized position in space and stop feeling returning force) (see Supplemental Video 1 for animation of the complete clustering process). During the ESD annealing process, the nodes are gradually and iteratively repositioned relative to all-against-all target distances. Each of the individual displacements is proportional to the deviation of target distances and thus represents a harmonic returning force. The harmonic spring constant for each bond can be set to uniform or differentially weighted in the following manner:

Equations 4a-c: Repositioning the nodes in search for the best match of the spatial pairwise displacement as an average of the displacements towards the target distance for each pair with the given node ($\vec{R}_i$ is the vector of the current position of the node in space, so the ratio of two positions divided by its length represents a unit vector directed from the current node i to its neighbor j, n=leaves on the tree):

Eq. 4a-Displacement with equal weighting

$$\vec{D}_i = \frac{\sum_{j=1}^{n} \frac{(\vec{R}_j - \vec{R}_i)(|\vec{R}_j - \vec{R}_i| - T_{ij})}{|\vec{R}_j - \vec{R}_i|}}{n}$$

Eq. 4b-Displacement weighted to favor close-neighbors

$$\vec{D}_i = \frac{\sum_{j=1}^{n} \frac{(\vec{R}_j - \vec{R}_i)(|\vec{R}_j - \vec{R}_i| - T_{ij})}{|\vec{R}_j - \vec{R}_i|^2}}{\sum_{j=1}^{n} \frac{1}{|\vec{R}_j - \vec{R}_i|}}$$

Eq. 4c-Displacement weighted to favor distant-neighbors

$$\vec{D}_i = \frac{\sum_{j=1}^{n} (\vec{R}_j - \vec{R}_i)(|\vec{R}_j - \vec{R}_i| - T_{ij})}{\sum_{j=1}^{n} |\vec{R}_j - \vec{R}_i|}$$

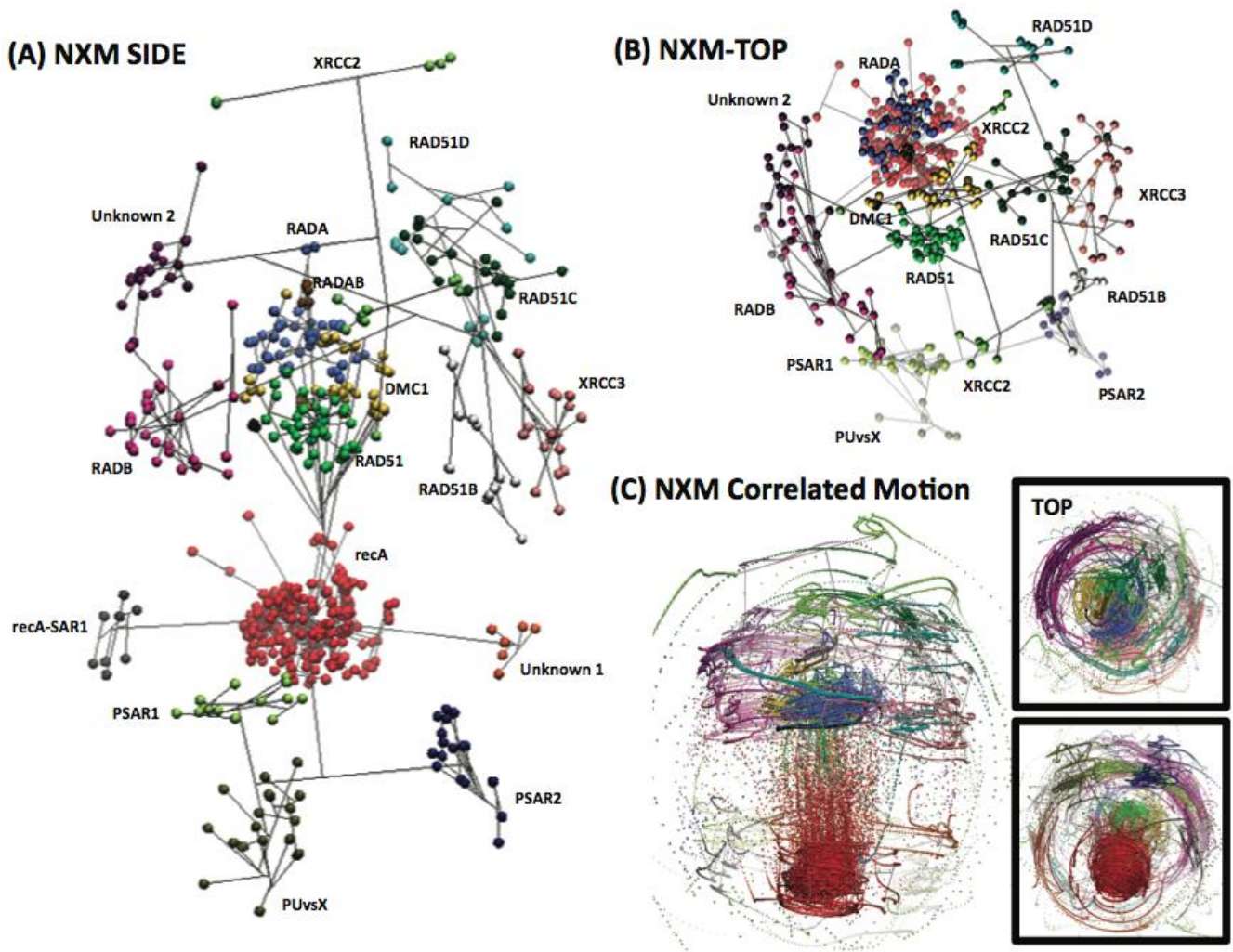

**Figure 8: Evolutionary Spatial Dynamics of the recA/RAD51 Superfamily.**
(A-B) Minimized tree of 633 recA/RAD51 sequences based directly on NXM similarity (rather than NXN Euclidian distances) shown from the side and top views, respectively. (C) This rendering visualizes every $10^{th}$ step of the whole 2000-step annealing trajectory as a series of colored nodes starting from an initial randomly distributed cloud. The further along the annealing trajectory, the closer the individual colored nodes are in series. Insets depict the top view and bottom view and are colored by subtype in the same manner as A-B.

For estimates on the spatial relationships between nodes one can use different weighting schemes for bonds described in Equation 4a-c: (i) equal weighting (i.e. uniform), (ii) close-neighbor weighting, or (iii) distant-neighbor weighting. We have tried all three weighting schemes, resulting in similar trees with reproducible features (data not shown). Nevertheless, there are a few important caveats. For example, both equal weighting and distant-neighbor weighting schemes tend to have a number of sequences randomly trapped behind a neighboring group. Conversely, the close-neighbor weighting scheme reproduces close- and long-range evolutionary relationships the best of the three schemes. Notably, the effectiveness of this close-neighbor weighting scheme might arise for a number of reasons, including: (i) the information for spatial positioning is progressively more reliable as similarity increases, while remote relatives bear the highest fraction of noise in the PHYRN-derived distance values, (ii) the proper spatial positioning of remote relatives is more precise when the local closer-relative structure is built first (i.e. there is a hierarchy to the tree assembly), and/or (iii) the spatial positioning of remote families integrates information from numerous pairwise contacts of comparable distances, while the local structure is dominated by just a few close-neighbor relationships. Therefore with respect to the latter case, the down-weighting of local relationships could disrupt proper spatial positioning of the close-neighbors, while up-weighting close-relationships might resolve them without negatively affecting the spatial positioning of remote relatives.

A key consideration for these simulations is that in a tree of n nodes there are $(n-1)^2$ distance constraints to satisfy and only $n*3$ spatial coordinates to vary. Thus, in a general case the exact solution for spatial positioning is impossible as there might be numerous local-optimum approximations. Therefore, the null hypothesis for this test is that there might be no clear or reproducible phylogenetic separation of groups if we start from a random distribution. In contrast, this algorithm leads to consistent groupings of individual clades. Figure 7A-C depicts a typical progression for the spatial dynamic clustering of 633 recA/RAD51 nodes (colored by subtype) over 1500 iterations. For clarity, connectivity is rendered in these panels only for close-neighbors while in the simulation every node is bonded to all other nodes. The structure, topology and trajectory files for representative minimized trees are available upon request and can be interrogated using open-source VMD software or any other molecular visualization software that can recognize PDB, PSF, and DCD formats. As mentioned, an important added value for ESD-based dynamics is that the entire clustering process can be visualized as a movie; hence, we have highlighted different aspects of this approach and various rendering options for the spatial dynamic models in Supporting Online Videos S1-2.

**Evolutionary Spatial Dynamics of recA/RAD51 Superfamily**

We explored a number of spatial dynamic simulations based on either the NXN similarity matrix (Figure 7D-F) or the NXM distance matrix (Figures 8-10). As previously described, the NXN distance matrix (utilized for PHYRN-NJ, see Figure 3-4) is constructed from an all-against-all Euclidian Distance of the NXM similarity matrix. Hence, pairwise comparisons in the NXN reflect the overall difference between all M-comparisons from the raw data. One important rationale for minimizing the spatial relationships in both NXM and NXN matrices is that they can provide visual landmarks that aid in conceptualizing the spatial changes of phylogenetic signal associated with the Euclidean Distance procedure itself.

Indeed, the dynamic motion and spatial relationships of nodes between this mathematical transformation is a form of correlated motion. Supplemental Video S2 depicts this dynamic transformation and shows that the largest changes between the two renderings are for the most divergent groups, with respect to the NXM metric. Intriguingly, we also observed another form of correlated motion, depicted in Figure 8C. Specifically, we observe shared clade-specific paths that nodes take during the search for minimized positioning in NXM simulations. In other words, this snapshot shows that even at the early stages of annealing, long before cladistic hierarchy can be observed, sequences more closely related to each other follow spatially correlated pathways towards their final positions. Taken together, these data suggest a potentially powerful theory - that relatedness can be quantitatively defined by how sequences move during the search even if they never find the global minima (i.e. a family of "imperfect solutions" is more informative than just one, even precisely found, minimum). This might reflect that the emergent behavior of the system experiences all sorts of small errors and noise, revealing the essential similarity despite the imperfection of the target distances matrix.

Importantly, both NXN and NXM simulations of 633 recA/RAD51 sequences produced a clear phylogenetic separation within 1500-3000 iterations (Figure 7 and 8, respectively). In data not shown, we repeated the minimization process multiple times starting with different: (i) initial random distributions and starting optimization points, (ii) scaling functions and (iii) weighting schemes. In all cases, the algorithm converges to similar shape with similar cladistic separation. Thus, the implication is that the models are robust and reproducible. Notably and in general, we did observe more distinctive phylogenetic boundaries in log scaled and close-neighbor weighted simulations compared with other settings. From the final converged 3D clusters, one can make a number of key observations: (i) that the spatial dynamic NXM minimizations largely accord with the heat maps in Figure 2, strongly suggesting that this is a reliable representation of PHYRN-derived phylogenetic signal, (ii) that the spatial dynamic NXN minimization are quite similar to the PHYRN-NJ topology and connectivity, implying that these multidimensional representations can be used to correlate evolutionary relationships with standard tree inference methods, and (iii) that the 3D distances of metagenomic clades relative to other groups lie within the range of distances between previously known recA/RAD51 groups. This further justifies the classification of these ancient groups in recA/RAD51 protein family and provides key inference on the basal evolutionary events.

Finally, as a separate test of classification accuracy, we challenged our NXM spatial clustering procedure for those thirteen RAD51 sequences from protist, fungi, plant, insect lineages and one metagenomic archaeal sequence (Unknown 2) that were unable to be unambiguously classified in PHYRN-NJ and caused major disruption to the branching pattern (see Table 1 and Methods). It is possible that some of these 14 sequences could be unrelated or incorrectly annotated relative to the rest of the recA/RAD51 sequences. While these sequences did lead to some changes in the 647 NXM minimized tree shape (data not shown) compared to the 633 NXM minimized tree (Figure 8A-B), the overall spatial relationships as a function of their putative group classification and phylogenetic signal were largely maintained. It is reasonable to consider that the observed distortion to the tree shape might be the result of: (i) the high level of divergence in these 14 sequences and/or (ii) a high degree of randomness in the similarity estimation.

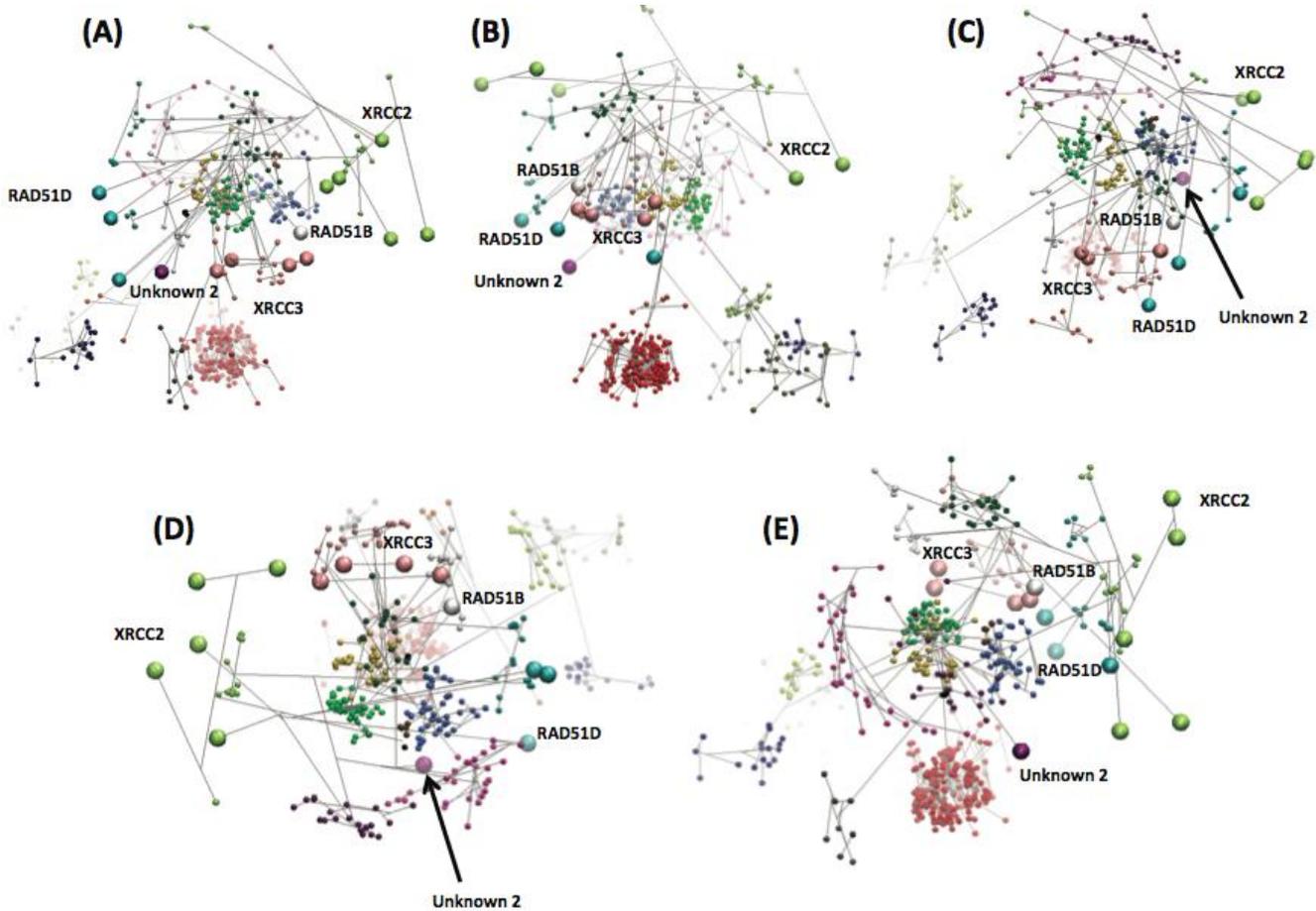

**Figure 9: Classification Test for Spatial Dynamic Simulations.** (A-B) Spatial positioning of 14 nodes (large spheres colored by putative group) over a set of pre-minimized 'fixed' NXM trees of 633 recA/RAD51 sequences. Each of five minimized 'fixed' trees of 633 recA/RAD51 sequences is based directly on NXM similarity (rather than NXN Euclidian distances).

Therefore, we attempted to position these 14 nodes over a set of pre-minimized 'fixed' NXM trees of 633 recA/RAD51 sequences (Figure 9). Thus, in this manner we might preserve the most reliable and consistent part of the phylogenetic information while annealing the new test sequences (i.e. in this case only the new sequences can move). We repeated the annealing procedure multiple times for a total of 100 independent simulations with different starting positions for the new sequences over five distinct minimized 'fixed' NXM trees (Figure 8A-E depicts individual replicate simulations for each of five 'fixed' tree shapes). While there are some exceptions, we observe that most of the sequences reliably find their putative group in this test. These data imply that the putative classification of these test sequences is largely correct. Future quantitative statistics will more probabilistically describe their behavior in conditions where: (i) all 647 sequences can dynamical reposition, or (ii) only 14 sequences can dynamically reposition around 633 'fixed' sequences.

As a comparison to the classification test of aforementioned 14 sequences, we applied a similar challenge to the randomized 633 recA/RAD51 sequences utilized for the calculation of the difference ratio in Figure 2B. Hence, these sequences are representative of unrelated sequences with a similar amino acid composition bias. We tried two different randomization tests of random spatial positioning for these sequences based on similarity to wild-type PSSM libraries: (i) spatial positioning of nodes representing the average composite scores of the randomized sequences (Figure 10A), or (ii) spatial positioning of individual random replicates for each of 633 recA/RAD51 sequences (Figure 10B). In addition, we subjected 633 viral RNA-dependent RNA-polymerases, a biological dataset unrelated to recA/RAD51 that also interacts with nucleotides, as a separate spatial dynamic test of random similarity (Figure 10C). In all cases, when these sequence sets are annealed over the fixed minimized tree, all of the randomized sequences position away from the tree on a slightly twisted "satellite" ring reminiscent of a Mobius strip. Notably, these random samples predominately accumulate in two areas approximating average equidistant positioning from all the 'fixed' sequences on the tree and without any noticeable cladistics separation. These results are particularly promising considering the fact that paralogous clades in the recA/RAD51 superfamily are in the 'Twilight-Zone" of sequence similarity (i.e. <25-30% pairwise sequence identity). Within this level of divergence it is often difficult to separate true relationships from random similarity with statistical confidence [18]. Future work in this regard will test additional biological datasets unrelated to the recA/RAD51 superfamily and qualitatively/quantitatively evaluate their behavior in dynamic models.

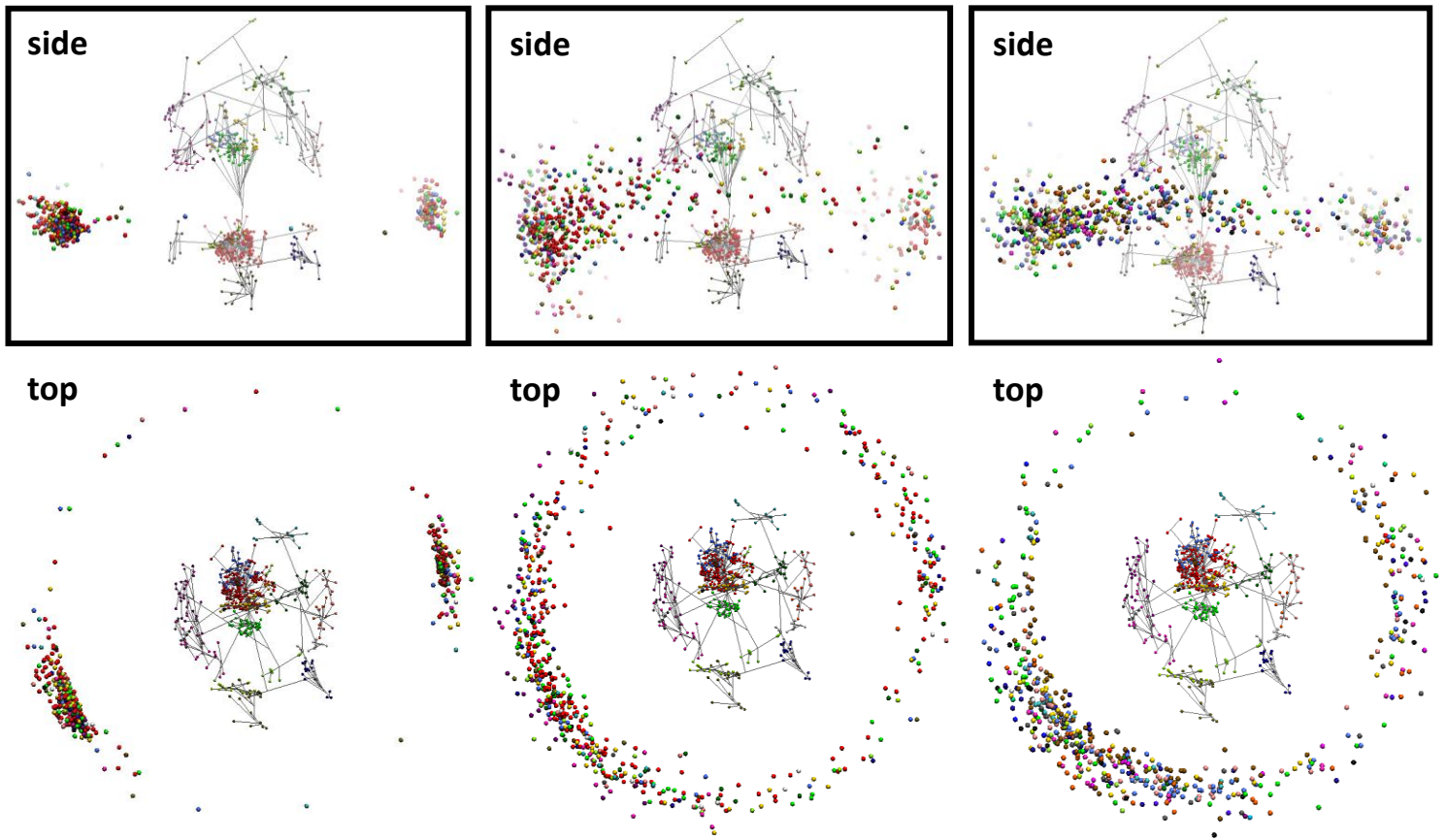

**Figure 10: Randomization Test for Evolutionary Spatial Dynamic Simulations.**
These snapshots represent NXM spatial dynamic simulations of randomized sequences around the 'fixed' tree of 633 recA/RAD51 sequences shown in Figure 8A-B. The target PHYRN-based distances utilized as inputs for the NXM-based ESD simulations were either (A) averaged composite scores for 633 recA/RAD51 sequences from 100 randomized trials, (B) composite scores for 633 recA/RAD51 sequences from individual randomized trials, or (C) composite scores for 633 viral RNA-dependent RNA-polymerases (RdRp) chosen at random from a dataset of 4206 RdRp. For all cases, side and top views are shown and the annealings were run for 2000 iterations.

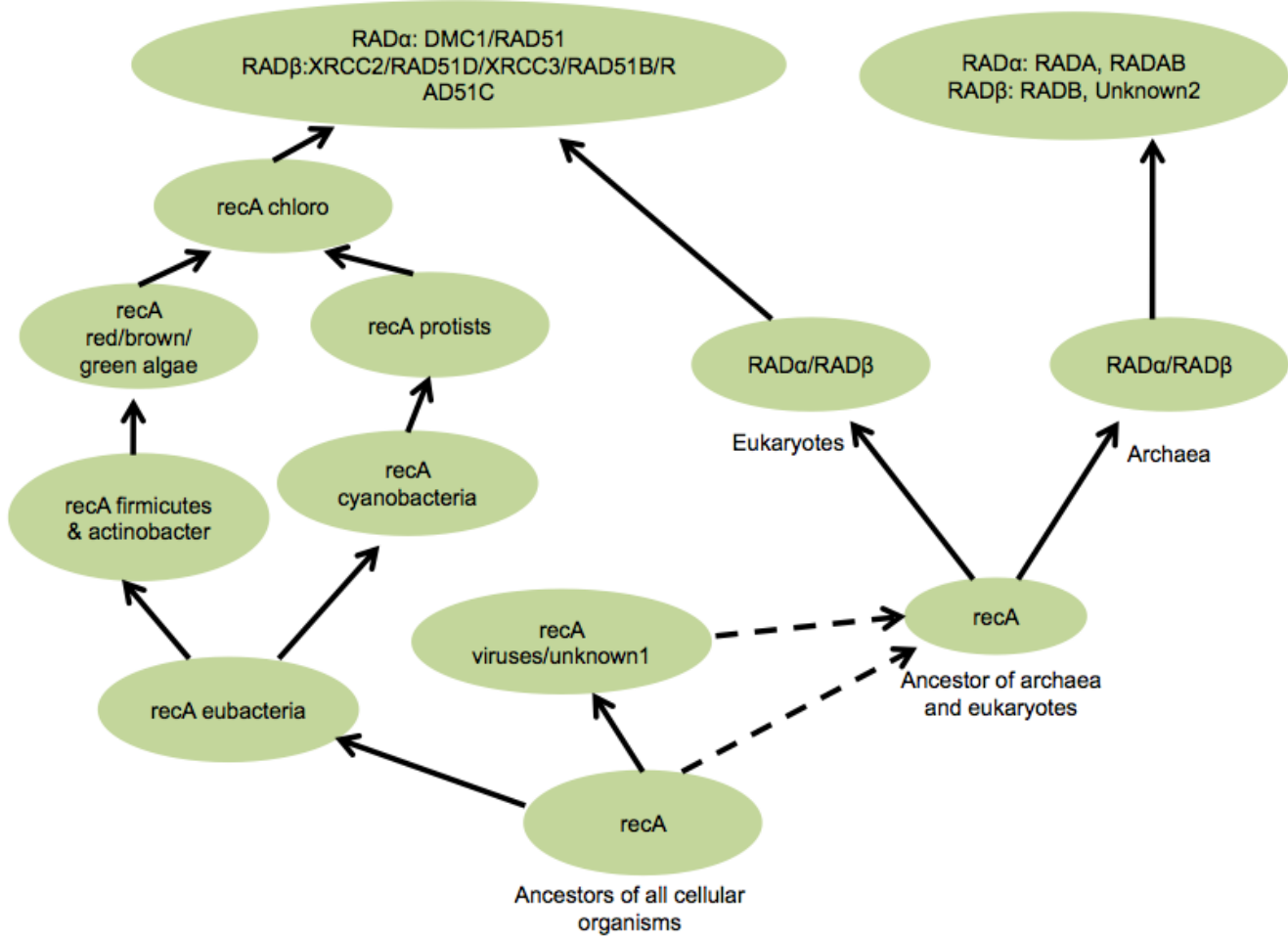

**Figure 11: Model of the Evolutionary History of the recA/RAD51 Superfamily**
Graphical representation of a model for evolution of recA/RAD51 family based on the phylogenetic trees obtained using PHYRN methodology.

## Discussion

We present a PHYRN-based phylogenetic inference for recA/RAD51, an ancient family of DNA repair proteins. Our combined results suggest that this phylogeny is more refined and resolved than previous reports considering our: (i) more comprehensive data set including newer, older and metagenomic sequences, (ii) refined metrics of distance estimation and simulation/visualization applications, and (iii) significant phylogenetic signal-to-noise ratio and larger statistical support across the entire landscape of protein representatives, despite the high levels of sequence divergence. Based on the PHYRN-derived phylogenetic trees and new spatial dynamic models, we propose a scenario for the evolution of recA/RAD51 family of proteins (Figure 11). In this model we make inferences on a number of key points, including: (i) the ancient origins of recA, (ii) differential rates of evolution for recA/RAD51 subfamilies, and (iii) the role(s) of endosymbiotic gene transfer events in the evolution of eukaryotic recA.

In our current model, the earliest recA evolved from a common ancestor of eubacteria and Unknown 1. With respect to recA, we infer multiple gene transfer events from cyanobacteria leading to evolution of chloroplast recA, in accordance with the origin of chloroplasts from cyanobacterial ancestors [33]. Intriguingly, based on the position and mutational rates of Unknown 1, our study corroborates the divergent nature of this group. Moreover, recA-SAR1 likely represents an intermediate group between Unknown 1 and known eubacterial clades (i.e. recA). Interestingly, the inferred rates of evolution in recA-SAR1 are very different from all other eubacterial clades, and are similar to evolutionary rates exhibited by members of Unknown 1.

It is well accepted that subsequent gene duplication events led to the diversification of ancient recA to RADα and RADβ in archaea and eukaryotes [16,17]. Our study also identifies an intermediate archaea group (RADAB) between RADA and RADB. Interestingly, both RADB and RADAB show monophyletic groups with members from the class euryarcheota, while RADA shows members from both major classes of archaea (i.e. crenoarcheota and euryarcheota). Within the RADA lineage, further gene duplications in protists presumably led to diversification of function into: (i) meiosis-specific DMC1 and (ii) RAD51, which have both somatic DNA repair and meiosis-specific genes. From the taxonomic diversity it is likely that DMC1 evolved in old alveolate members. Moreover, it is possible that DMC1 in higher eukaryotes attained a more specialized meiosis-specific role through multiple loss of functional mutations overtime. In the RADB lineage, we propose, in contrast to Wu *et al.*[17], that Unknown 2 attains a position closer to RADB. Given that both these groups are archaea-specific this positioning is more plausible. Further, we infer at least two gene duplications in archaea: eukaryotic RAD51D, XRCC3, RAD51B and RAD51C evolved as results of the first duplication while eukaryotic XRCC2 might have evolved in a second gene duplication event in RADB lineage.

Taken together, this study makes a number of contributive advances: (i) we present further validation of PHYRN-based inference in an ancient protein family with variable rates, (ii) we provide a dynamic and interactive simulation/visualization/analytical tool for 3D representation of evolutionary distances, and (iii) we derive a refined model of recA/RAD51 evolution. Moreover, our spatial dynamic simulations suggest that different scaling and scoring functions can make implementation of PHYRN-measures even more resistant to sequences with low information, conflicts in distance, and/or variable rates of change. Future work will be focused on these types of optimization schemes for PHYRN and the utility of evolutionary spatial dynamics in this regard. Finally, we echo the sentiments of Wu *et al.*[17] in that we also believe that annotation of more metagenomic recA sequences and their inclusion in the phylogenetic inference is essential for a deeper and more refined understanding of recA/RAD51 phylogeny and endosymbiotic transfer events in general.

**Acknowledgements**

This work was supported by the Searle Young Investigators Award and start-up money from UC Davis (RLP), and The National Institutes of Health R01 GM087410-01 (RLP, DVR). This project was also funded by a Fellowship from the Eberly College of Sciences and the Huck Institutes of the Life Sciences (DVR) and a grant with the Pennsylvania Department of Health using Tobacco Settlement Funds (DVR). The Department of Health specifically disclaims responsibility for any analyses, interpretations or conclusions. We would like to thank Maia Rabaa, Ngai Lam Ho, Natasha Shah, and Alyssa Thunen for their help and support during the project, as well as Jason Holmes at The Pennsylvania State University CAC center for technical assistance. We would like to thank Dr. Robert E. Rothe, Euclid O. Alexandria, Barbara Van Rossum, Jim White, and Benoit Mandlebrot for creative dialogue.

**Supporting Spatial Dynamic Video-1: Typical Progression of a NXN Spatial Dynamic Minimization (https://www.dropbox.com/s/v8lz60rc0cijoya/movie.avi)**

In our current paradigm of evolutionary spatial dynamics, each sequence is represented as a node that is connected to all other nodes by different types of bonds (i.e. forces proportional or inverse to deviation) that drive network towards PHYRN-derived pairwise distances and are indicative of different weighting schemes. This NXN spatial dynamic movie shows: (i) the starting randomized conformation of the nodes representing 633 sequences from the recA/RAD51 Recombinase superfamily family, (ii) spatial clustering of the nodes based on all-to-all pairwise similarity distances, (iii) visual overlapping with the independently derived PHYRN tree connections (NJ method based on the pairwise distances matrix, not used for spatial clustering), and (iv) rendering of surfaces enclosing the spatial clusters based on the spatial density of the nodes.

**Supporting Spatial Dynamic Video-2 Transformation between Typical NXM and NXN Spatial Dynamic Minimizations (https://www.dropbox.com/s/avaoy3od5aj17c2/log-nXm_linear-nXn_trees-interpolation.avi)**

The movie highlights the transformation between the minimized NXN- and NXM-based trees (see Figure 7-8, respectively). This was accomplished using linear interpolation between the positions in the annealed trees, and aligned to minimize RMSD between the two structures. Labels for specific recA/RAD51 clades are as follows: DMC1, 1; RAD51/DMC1 Giardia, 2; recA-SAR1, 3; Unknown-1, 4; PSAR1, 5; PUvsX, 6; PSAR2, 7; Unknown-2, 8; RAD51, 9; RAD51B, 10;RAD51C, 11; RAD51D, 12; RADA, 13; RADAB, 14; RADB, 15; recA, 16; XRCC2, 17; XRCC3, 18.

# References

1. Thompson LH, Schild D (2001) Homologous recombinational repair of DNA ensures mammalian chromosome stability. Mutation research 477: 131-153.
2. van den Bosch M, Lohman PH, Pastink A (2002) DNA double-strand break repair by homologous recombination. Biological chemistry 383: 873-892.
3. Thacker J (2005) The RAD51 gene family, genetic instability and cancer. Cancer letters 219: 125-135.
4. Bishop DK, Park D, Xu L, Kleckner N (1992) DMC1: a meiosis-specific yeast homolog of E. coli recA required for recombination, synaptonemal complex formation, and cell cycle progression. Cell 69: 439-456.
5. Shinohara A, Ogawa H, Ogawa T (1992) Rad51 protein involved in repair and recombination in S. cerevisiae is a RecA-like protein. Cell 69: 457-470.
6. Eisen JA (1995) The RecA protein as a model molecule for molecular systematic studies of bacteria: comparison of trees of RecAs and 16S rRNAs from the same species. Journal of molecular evolution 41: 1105-1123.
7. DiRuggiero J, Brown JR, Bogert AP, Robb FT (1999) DNA repair systems in archaea: mementos from the last universal common ancestor? Journal of molecular evolution 49: 474-484.
8. Komori K, Miyata T, DiRuggiero J, Holley-Shanks R, Hayashi I, et al. (2000) Both RadA and RadB are involved in homologous recombination in Pyrococcus furiosus. The Journal of biological chemistry 275: 33782-33790.
9. Golubovskaya IN, Hamant O, Timofejeva L, Wang CJ, Braun D, et al. (2006) Alleles of afd1 dissect REC8 functions during meiotic prophase I. Journal of cell science 119: 3306-3315.
10. Lovett ST (1994) Sequence of the RAD55 gene of Saccharomyces cerevisiae: similarity of RAD55 to prokaryotic RecA and other RecA-like proteins. Gene 142: 103-106.
11. Game JC (1993) DNA double-strand breaks and the RAD50-RAD57 genes in Saccharomyces. Seminars in cancer biology 4: 73-83.
12. Gaasbeek EJ, van der Wal FJ, van Putten JP, de Boer P, van der Graaf-van Bloois L, et al. (2009) Functional characterization of excision repair and RecA-dependent recombinational DNA repair in Campylobacter jejuni. Journal of bacteriology 191: 3785-3793.
13. Tsuzuki T, Fujii Y, Sakumi K, Tominaga Y, Nakao K, et al. (1996) Targeted disruption of the Rad51 gene leads to lethality in embryonic mice. Proceedings of the National Academy of Sciences of the United States of America 93: 6236-6240.
14. Li W, Ma H (2006) Double-stranded DNA breaks and gene functions in recombination and meiosis. Cell research 16: 402-412.
15. Stassen NY, Logsdon JM, Jr., Vora GJ, Offenberg HH, Palmer JD, et al. (1997) Isolation and characterization of rad51 orthologs from Coprinus cinereus and Lycopersicon esculentum, and phylogenetic analysis of eukaryotic recA homologs. Current genetics 31: 144-157.
16. Lin Z, Kong H, Nei M, Ma H (2006) Origins and evolution of the recA/RAD51 gene family: evidence for ancient gene duplication and endosymbiotic gene transfer. Proceedings of the National Academy of Sciences of the United States of America 103: 10328-10333.
17. Wu D, Wu M, Halpern A, Rusch DB, Yooseph S, et al. (2011) Stalking the fourth domain in metagenomic data: searching for, discovering, and interpreting novel, deep branches in marker gene phylogenetic trees. PLoS One 6: e18011.
18. Bhardwaj G, Ko KD, Hong Y, Zhang Z, Ho NL, et al. (2012) PHYRN: A Robust Method for Phylogenetic Analysis of Highly Divergent Sequences. PLoS One 7: e34261.
19. Sober E (1983) Parsimony in Systematics: Philosophical Issues. Annual Review of Ecology and Systematics 14: 335-357.
20. Stamatakis A, Hoover P, Rougemont J (2008) A rapid bootstrap algorithm for the RAxML Web servers. Systematic biology 57: 758-771.
21. Huelsenbeck JP, Ronquist F (2001) MRBAYES: Bayesian inference of phylogenetic trees. Bioinformatics 17: 754-755.
22. Marchler-Bauer A, Lu S, Anderson JB, Chitsaz F, Derbyshire MK, et al. (2011) CDD: a Conserved Domain Database for the functional annotation of proteins. Nucleic acids research 39: D225-229.
23. Altschul SF, Madden TL, Schaffer AA, Zhang J, Zhang Z, et al. (1997) Gapped BLAST and PSI-BLAST: a new generation of protein database search programs. Nucleic acids research 25: 3389-3402.
24. Eisen MB, Spellman PT, Brown PO, Botstein D (1998) Cluster analysis and display of genome-wide expression patterns. Proceedings of the National Academy of Sciences of the United States of America 95: 14863-14868.
25. Saitou N, Nei M (1987) The neighbor-joining method: a new method for reconstructing phylogenetic trees. Molecular biology and evolution 4: 406-425.
26. Nei M, Kumar S, Takahashi K (1998) The optimization principle in phylogenetic analysis tends to give incorrect topologies when the number of nucleotides or amino acids used is small. Proceedings of the National Academy of Sciences of the United States of America 95: 12390-12397.
27. Desper R, Gascuel O (2002) Fast and accurate phylogeny reconstruction algorithms based on the minimum-evolution principle. J Comput Biol 9: 687-705.
28. Tamura K, Peterson D, Peterson N, Stecher G, Nei M, et al. (2011) MEGA5: molecular evolutionary genetics analysis using maximum likelihood, evolutionary distance, and maximum parsimony methods. Molecular biology and evolution 28: 2731-2739.
29. Edgar RC (2004) MUSCLE: multiple sequence alignment with high accuracy and high throughput. Nucleic acids research 32: 1792-1797.
30. Felsenstein J (1993) PHYLIP (Phylogeny Inference Package) Department of Genetics, University of Washington, Seattle.: Distributed by the author.
31. Felsenstein J (1989) PHYLIP -- Phylogeny Inference Package Cladistics Version 3.2: 164-166.
32. Jones DT, Taylor WR, Thornton JM (1992) The rapid generation of mutation data matrices from protein sequences. Computer applications in the biosciences : CABIOS 8: 275-282.
33. Price DC, Chan CX, Yoon HS, Yang EC, Qiu H, et al. (2012) Cyanophora paradoxa genome elucidates origin of photosynthesis in algae and plants. Science 335: 843-847.
34. Dunn CW, Hejnol A, Matus DQ, Pang K, Browne WE, et al. (2008) Broad phylogenomic sampling improves resolution of the animal tree of life. Nature 452: 745-749.
35. Smith SA, Beaulieu JM, Donoghue MJ (2009) Mega-phylogeny approach for comparative biology: an alternative to supertree and supermatrix approaches. BMC Evol Biol 9: 37.
36. Kim N, Lee C (2007) Three-Dimensional Phylogeny Explorer: distinguishing paralogs, lateral transfer, and violation of "molecular clock" assumption with 3D visualization. BMC Bioinformatics 8: 213.
37. Hughes T, Hyun Y, Liberles DA (2004) Visualising very large phylogenetic trees in three dimensional hyperbolic space. BMC Bioinformatics 5: 48.
38. Huson DH, Bryant D (2006) Application of phylogenetic networks in evolutionary studies. Molecular biology and evolution 23: 254-267.
39. Woolley SM, Posada D, Crandall KA (2008) A comparison of phylogenetic network methods using computer simulation. PLoS One 3: e1913.
40. Humphrey W, Dalke A, Schulten K (1996) VMD: visual molecular dynamics. J Mol Graph 14: 33-38, 27-38.
41. Gilbert DR, Schroeder M, van Helden J (2000) Interactive visualization and exploration of relationships between biological objects. Trends Biotechnol 18: 487-494.
42. Michael Schroedera, , David Gilberta, , Jacques van Heldenb, , Penny Noya, (2001) Approaches to visualisation in bioinformatics: from dendrograms to Space Explorer. Information Sciences 139: 19-57.
43. Atkinson HJ, Morris JH, Ferrin TE, Babbitt PC (2009) Using sequence similarity networks for visualization of relationships across diverse protein superfamilies. PLoS One 4: e4345.